\documentstyle[12pt,aasms4]{article}

\slugcomment{Submitted to Astrophysical Journal Letters}

\lefthead{Aretxaga et al.}
\righthead{Seyfert~1 mutation of the classical Seyfert 2 nucleus NGC~7582}

\newcommand{\Ha}{\mbox{H{\small $\alpha$}}}
\newcommand{\ldo}[1]{\hbox{$\lambda$#1 \AA}}
\newcommand{\ldoldo}[2]{\hbox{$\lambda \lambda$#1/#2 \AA}}
\newcommand{\univel}{\hbox{km s$^{-1}$}}
\newcommand{\uniHo}{\hbox{km s$^{-1}$ Mpc$^{-1}$}}
\newcommand{\Hb}{\mbox{H{\small $\beta$}}}
\newcommand{\Ho}{\hbox{H$_{\mbox{\scriptsize 0}} = 50$~\uniHo \ }}
\newcommand{\lsim}{\mathrel{\lower2.5pt\vbox{\lineskip=0pt\baselineskip=0pt
           \hbox{$<$}\hbox{$\sim$}}}}
\newcommand{\gsim}{\mathrel{\lower2.5pt\vbox{\lineskip=0pt\baselineskip=0pt
           \hbox{$>$}\hbox{$\sim$}}}}

\begin{document}

\title{Seyfert~1 mutation of the classical Seyfert 2 nucleus 
NGC~7582\altaffilmark{1}}

\author{Itziar Aretxaga}
\affil{Instituto Nacional de Astrof\'{\i}sica, \'Optica y 
Electr\'onica,\\ Apdo. Postal 25 y 216, 72000 Puebla, Pue., Mexico\\
E-mail: itziar@inaoep.mx}

\author{B. Joguet}
\affil{European Southern Observatory,\\ Alonso de Cordova 3107,
Vitacura, Casilla 19001, Santiago 19, Chile\\
Institut d'Astrophysique de Paris, 98 bis Boulevard Arago,
F-75014 Paris, France\\
E-mail:bjoguet@eso.org}

\author{D. Kunth}
\affil{Institut d'Astrophysique de Paris, 98 bis Boulevard Arago,
F-75014 Paris, France\\
E-mail: kunth@iap.fr}

\author{J. Melnick}
\affil{European Southern Observatory,\\ Alonso de Cordova 3107,
Vitacura, Casilla 19001, Santiago 19, Chile\\
E-mail: jmelnick@eso.org}

\and

\author{R. J. Terlevich\altaffilmark{2}}
\affil{Institute of Astronomy, Madingley Road, 
Cambridge CB3 0HA, U.K.\\
E-mail: rjt@ast.cam.ac.uk}


\altaffiltext{1}{
Based on observations collected at the European Southern
Observatory at La Silla (Chile)}
\altaffiltext{2}{Visiting Professor at 
INAOE, Puebla, Mexico}


\begin{abstract}
We report the transition
towards a type~1 Seyfert experienced by the classical
type~2 Seyfert nucleus in NGC~7582. The transition, found at most 20 days
from its maximum peak, presents a unique opportunity to study 
these rare events in detail. 
At  maximum the \Ha\ line width  is of about  12000~\univel. 
We examine three scenarios 
that could potentially 
explain the transition: capture of a star by a supermassive
black hole, a reddening change in the surrounding torus, and the
radiative onset of a type IIn supernova
exploding in a compact nuclear/circumnuclear starburst.
\end{abstract}




%

\section{Introduction}

	The last decade has seen a significant advance in our way of
understanding the phenomenology of various classes of Active Galactic
Nuclei (AGN) under the concept of orientation (see Antonucci 1993 for a 
review).
Unified schemes postulate
the existence of a dusty obscuring torus that hides both
the Broad Line Region (BLR) and the continuum emitting zone
when seeing the nuclear region of an AGN edge-on. Seyfert 2 and 1 nuclei
would then be exactly the same kinds of objects but viewed from different
angles, such that the line of sight crosses or not the dusty torus.
This configuration can explain several observables of Seyfert~2 nuclei,
like the detection of broad lines in polarized
light at optical wavelengths
(Antonucci \& Miller 1985, Miller \& Goodrich 1990), which are 
directly detected in near-IR light (Goodrich, Veilleux \& Hill 1994),
the presence of kpc scale ionizing cones 
(Pogge 1988,1989, Tadhunter \& Tsvetanov 1989),
and the steepening and excesses of far-IR colours
(Maiolino et al. 1995, Spinoglio et al. 1995, P\'erez Garc\'{\i}a,
Rodr\'{\i}guez-Espinosa \& Santolaya Rey 
1998).

	A problem still arises when analyzing in detail where the blue 
continuum observed in Seyfert~2 nuclei comes from. Classically,
this was regarded to be light scattered by dust or hot electrons in the
ionized region illuminated by the nucleus. However, the different
polarization levels of broad lines and continuum (e.g. 
Miller \& Goodrich 1990, Tran, Miller \& Kay 1992) and the
absence of residual 
broad lines in directly
received optical light argue in favour of the existence of a second 
continuum source, most probably a region of star formation that 
surrounds the
dusty torus (Goodrich 1989a, Cid Fernandes \& Terlevich 1995; Heckman et 
al. 1995). Starbursts
in the nuclei of Seyfert~2 galaxies have indeed been detected
through near-IR photospheric absorption features characteristic of 
red supergiant stars (Terlevich, D\'{\i}az \& Terlevich 1990;
Oliva et al. 1995). Some of these 
dominate the observed continuum from the near-UV to the near-IR,
and have resolved sizes of a few hundreds of pc (Heckman et al. 1997; 
Colina et al. 1997; Gonz\'alez Delgado et al. 1998).

NGC~7582 is a classical Seyfert~2 nucleus, as indicated by the
relatively high excitation measured from the
[O~II]\ldo{3727}/[OIII]\ldo{5007} ratio plus the simultaneous presence 
of strong [Ne~V]\ldo{3425},
He~II\ldo{4686}, [O~I]\ldo{6300} and [N~II]\ldoldo{6548}{6584} lines.
Many of the nuclear properties in NGC~7582 also support a
unified scheme.
A sharp-edged [O~III] outflow in the form of a cone is observed
(Morris et al. 1985, Storchi-Bergmann \& Bonatto  1991).
Optical spectropolarimetry does not reveal any hidden BLR, but since the
far-IR colours $60\mu\mbox{m} - 25\mu$m are very red, the absence has been 
taken as support for an edge-on thick torus able to block even the light
scattered towards the
observer (Heisler, Lumsden \& Bailey 1997). Indeed
a large column density of neutral H also blocks the hard X-rays, 
implying a large obscuration (Warwick et al. 1993). 

The presence of stars in the nucleus is now firmly established.
Morris et al. (1985) found 
a steep gradient of \Ha\  perpendicular to the [O~III] cone,
which they interpret as a
1~kpc disk of H~II regions oriented at $60^o$ from the plane of the galaxy 
(see their Figure~10). 
The CO absorption lines and large near-IR light-to-mass ratio are
similar to those of H~II galaxies and a factor of 5 larger 
than those of normal galaxies, indicating that red supergiants dominate
the light of the inner 200~pc at those 
wavelengths (Oliva et al. 1995). 

	This letter analyzes new spectroscopic and surface photometry
data that show that NGC~7582 has experienced a transition to a Seyfert~1
stage. This transition is a challenge to reconcile with unification
squemes since the IR to X-ray data indicate that we are viewing
the nucleus through a large column of obscuring material.

\section{Data acquisition and analysis}

 Three spectra of NGC~7582 were obtained at 
ESO-La Silla 
with the Danish 1.54m and 
ESO 3.6m telescopes from 1998 July 11th
to October 21st. The journal of
observations in Table~1 contains the relevant information
regarding set-ups and exposure times. 
The slit was always centered on the nucleus of the galaxy and oriented at
24$^o$ PA. The total integration time per night was split in two or three 
integrations in order 
to be able to remove the effect of cosmic rays in the co-added final spectrum.
Standard stars were observed all nights, except on October 6th. 
The frames have a
scale of 0.39 arcsec/pixel with the 1.54m telescope and 0.15 arcsec/pixel 
with the  3.6m.

 The data were reduced using the {\sc iraf} 
software package in a standard way. 
The CCD frames were first bias subtracted and then flat-field corrected.
Wavelength calibration with He+Ar lamps
and flux calibration with the corresponding stars was then performed, and 
the sky was subtracted using the outermost parts of the slit.

  The spectra were internally flux-calibrated to the same relative scale 
using [N~II]\ldo{6583} and, independently, also [O~III]\ldo{5007}, under 
the standard assumption that the flux of 
the narrow forbidden lines of AGN does not vary on such short time-scales 
(e.g. Korista et al. 1995).

 Two $R$-band images of 30 and 90~s 
were also obtained on July 11th and October 21st
with the same telescope configuration. The reduction involved
bias subtraction and flat-fielding. Calibration with a photometric
standard star was obtained relative to the field stars of the frames.

Figure~1 displays spectra that we extracted binning pixels
along the spatial direction over 2 arcsec,
which corresponds to 150~pc at 
the recession velocity of the galaxy $v_r=1575$~\univel\ 
(de Vaucouleurs et al. 1991) for
\Ho. The top panel shows a spectrum of NGC~7582 taken 
well before the transition (Cid Fernandes, Storchi-Bergmann \& Schmitt 1998).
This remained unchanged as a Seyfert~2 spectrum until at least 1998 June 20th 
(Halpern, Kay \& Leight 1998). 
Note that the spectrum of October 6th is not flux-calibrated.

The spectrum of July 11th clearly shows broad \Ha, \Hb, 
Na~I and Fe~II emission, absent in the previous record.
The broad component of the Balmer emission lines is well represented
by a single gaussian of FWHM=12400~\univel\
blue-shifted by 1980~\univel\ with respect to the narrow 
FWHM=260~\univel\ 
lines. 
By October 6th, the broad \Hb\ has disappeared, but broad Fe~II, Na~I and 
a prominent
\Ha\ of FWHM=7870 \univel\ is still present, redshifted by 1570~\univel\
with respect to the narrow lines. The October 21st spectrum 
shows a further decline of the broad \Ha\ line to FWHM=6660 \univel.
It should be noted that although the broad \Hb\ line is not apparent
in the spectra after October 6th, when the bulge population is 
subtracted, a very weak broad line is indeed present.

  The continuum 
flux of the nucleus at \ldo{6260} decreases by a 50\% between July 11th 
and October 
6th, and increases by a 30\% between October 6th and 21st. 

In order to check if the transition is really nuclear we have calculated 
the astrometry of the nucleus relative to field stars in our
$R$-band images obtained 
after the transition, and in an 
archival image taken on 1995 June 16th
(before the transition) 
with the Wide Field Planetary Camera~2 (WFPC2) of the 
Hubble Space Telescope (HST).
We find that the centroid of the nucleus
is at the same position, inside an error box of 0.16~arcsec, i.e.
12~pc.




\section{Discussion}

Figure~1 shows that shortly before July 11th the nucleus of
NGC~7582 developed broad permitted lines characteristic
of a Seyfert~1 nucleus ( Joguet, Kunth \& Terlevich 1998). 

While variability is a common characteristic of 
Seyfert~1 nuclei and QSOs, which are
known to undergo prominent variations (e.g. Peterson 1993) that 
lead even to the temporal disappearance of broad lines
(e.g. Antonucci \& Cohen 1983, Penston \& P\'erez 1984, Alloin  et al. 1986), 
it is actually a rarity among Seyfert~2 and 
{\sc liner}s.
Apart from the transition experienced by NGC~7582, reported here, 
three other classical Seyfert~2 nuclei and one 
{\sc liner} developed
broad lines in the past: Mrk~6 (Khachikian \& Weedman 1971), 
Mrk~993 (Tran, Osterbrick \& Marel 1992), Mrk~1018 (Cohen et al. 1986) and 
NGC~1097 (Storchi-Bergmann, Baldwin \& Wilson 1993). 
The nucleus of NGC~7582 is however 
the only one caught {\it in fraganti} at such 
an early 
stage in the transition.

  In some cases the sudden development of broad lines has 
been interpreted in the 
literature as transients associated with the disruption and 
capture of stars by a 
supermassive black hole. The frequency of such events in a normal spiral 
galaxy is
predicted to be $10^{-4}$ yr$^{-1}$ (Rees 1988). 
The number of narrow-lined AGN
registered in the V\'eron-Cetty \& V\'eron catalogue (1998) is 701, 
including those with reported obscured or scattered broad-lines.
The  probability of five of them undergoing a capture phenomenon
in the last 30~years is  4\%, assuming a Poissonian
temporal
distribution of the captures and a continuous monitoring or long-lasting
events. In the case of NGC~7582
this could be possible
if the idea of the dusty torus around the nucleus 
is dropped. A torus would otherwise block all the optical light coming from
the surroundings of the black hole. 

  
  As a matter of fact two of the transition  cases mentioned above
(Mrk~993  and Mrk~1018)
have been proposed to be explained 
in the framework of unification by patchy dusty tori that 
suddenly let the nuclei be directly seen  through
regions with small covering factors or light obscuration. Goodrich
(1989b, 1995) finds evidence for a very weak, previously unnoticed, 
broad \Ha\ line before the transitions to Seyfert~1 stages in these two nuclei
and argues that the flux
changes derived from continuum bands and broad lines
at different wavelengths are consistent with the simple hypothesis of 
a change in reddening.
In the case of NGC~7582 the BLR reddening before and after the transition
cannot be checked since
we find no evidence of a weak broad line before the
transition. The continuum change measured at \ldo{5026} and
\ldo{6260} is however inconsistent with a reddening change given by 
a local reddening law: the fluxes at \ldo{5026} 
of  the 1994 January 6th and 1998 July 11th spectra, after 
bulge subtraction, imply 
$\Delta E_{B-V} \approx 0.3$~mag, and at \ldo{6260} 
$\Delta E_{B-V} \approx 1.3$~mag. Some residual stellar light must still
be present here since young stellar populations are 
located near the nucleus. Schmitt, Storchi-Bergmann \& Cid Fernandes (1998) 
account these to be 
at least a 6\% of the total nuclear light.
If we assume a continuum shape typical of starbursts,
$f_\nu \propto \nu^{\beta}$ with $\beta \approx 0 - 1$ at an age 
$\lsim 10^7$~yr
(e.g. Leitherer et al. 1999), we still 
have reddening changes ranging 
$\Delta E_{B-V} \approx 1.1$ to 2.3~mag at \ldo{5026} and 
$\Delta E_{B-V} \approx 3.3$ to 3.5~mag at \ldo{6260}, inconsistent 
with one another. Furthermore, it would be difficult to understand 
how the BLR light 
has managed to traverse a  torus  with about
200~mag of extinction ($A_V$), as derived from 
X-rays absorptions. However,
the column density of absorbing material has been reported to change 
even a factor of 3 in the last 13 years (Warwick et al. 1993,
Xue et al. 1999). X-ray data will shed new
light into  the 
possibility of a reddening change in the torus. 
Possible caveats of the  optical estimations are, one one hand, 
that these are based on the local
reddening law  and, on the other, 
that the contamination of the young nuclear/circumnuclear 
stellar component has been removed through modeling in the absence of 
high spatial resolution and/or a wider spectral coverage.

A more fundamental concern in this analysis 
is whether the BLR and the continuum light will actually
be crossing the same obscuring regions of the torus. 
In principle, a model could be built in which 
the BLR light might pass through clearance zones not reached by 
the continuum light. This, however, seems rather {\it ad hoc}.

An alternative possibility to explain the transition can be found
resourcing to  phenomena that might occur around the torus, and which are not 
necessarily related to the 
central engine of the AGN. The nuclear/circumnuclear starburst, detected at 
a radius $r<100$~pc (Oliva et al. 1995), is a possible  source since
it must produce supernova (SN) explosions. 

From an age 10 -- 60~Myr, 
a starburst sustains a SN~rate directly proportional 
to the blue light emitted by the stars in the cluster
(Aretxaga \& Terlevich 1994).
 The luminosity of the inner 3~arcsec of NGC~7582 before the transition
is $V = 15$~mag (Kotilainen, Ward \& Williger 1993). The luminosity profile
at this stage 
is not very peaked, as shown in HST images (Malkan, Gorjian \& Tam 1998),
where the luminosity of the nucleus changes by 3~mag when using
apertures ranging from 0.5 to 0.1~arcsec. 
The stellar populations of the nucleus at optical wavelengths 
are best represented by old 
ages ($t>100$~Myr) with a 6 to 12\% contamination  by a reddened 
$E_{B-V}\approx 0.6$~mag
starburst + featureless continuum (Schmitt et al. 1998).
Therefore, the intrinsic luminosity of the starburst nucleus is about 
$M_B\approx -17$~mag and
the SN rate would then be 
$\nu\approx0.02$~yr$^{-1}$. 
If all the SN explosions generate broad lines, there 
is a 33\% probability of having detected
a transition to a Seyfert~1 stage in the last 30~yr.

A typical SN can increase the luminosity of the host cluster by about 2~mag.
Classical type~II SN, however, show spectral features, like P-Cygni
profiles, which do not compare with the observed gaussian profiles
of the Balmer lines in NGC~7582. The only possible exception could be SN~1983K,
which showed prominent broad emission lines at maximum and a typical plateau
evolution of about 100~days (Phillips et al. 1990).

On the other hand, 
since the discovery a new kind of SN with
spectral features similar to those of Seyfert~1 nuclei 
(Filippenko 1989),
there has been a wide record of new events in SN catalogues, and
a body of studies that highlight their exceptional characteristics. 
The new SN group, named SN~IIn (Schlegel 1990, Filippenko 1997), show variable 
broad emission
lines of up to FWHM$\approx 20000$~\univel, sitting below prominent
narrow emission lines (hence the `n'),
and a light decay much flatter than typical SN.
The absence of the characteristic
P-Cygni profiles of expanding atmospheres throughout their evolution 
is remarkable. 
Some of these SN~IIn are exceptionally bright at
radio, optical and X-ray wavelengths, with total radiated energies
of several times $10^{51}$~erg in less than a decade  (Aretxaga et al. 1999), 
i.e. two orders of magnitude more than typical SN events.
These properties have been interpreted in the light of
quick reprocessing of the kinetical energy released in the explosion 
by a dense circumstellar medium (Chugai 1991, Terlevich et al. 1992),
and thus explain the phenomenon as a young and compact supernova remnant 
(cSNR) rather than a SN.

  There is little doubt that if a SN~IIn explodes in the center of a normal 
galaxy, the nucleus would be classified as a Seyfert~1, while 
the prominent broad lines remain visible. In fact, there has been
a succession of theoretical works that explain the phenomenology 
of lines and continuum at UV to near-IR 
wavelengths in Seyfert~1 nuclei in terms of a starburst that undergoes
SN~IIn explosions 
(e.g. Terlevich et al. 1992, 1995, Aretxaga \& Terlevich 1994).

Figure~2 plots the evolution of the 
\Ha\ line width and luminosity as a function of time compared with
the values found in one of the best followed-up SN~IIn SN~1988Z (data from
Aretxaga et al. 1999), and the \Hb\ luminosity evolution of the
Seyfert~1 NGC~5548 (data from Korista
et al. 1995). The time axis for the comparison is uncertain. We have opted to
match the light curves of SN~1988Z, NGC~7582 and NGC~5548 at the maximum 
light. This also matches the minimum 
in the light evolution of NGC~5548 as the onset of the
'elementary unit of variation' in Seyfert~1 nuclei proposed by
Cid Fernandes, Terlevich \& Aretxaga (1997). 
All fluxes have been normalized to the maximum value. 
The line-width evolution of NGC~7582 closely resembles the early evolution
of SN~1988Z, but the early evolution of the line intensity seems much steeper 
than that experienced by SN~1988Z or NGC~5548. 

A close 
multi-wavelength monitoring of this object is thus required if we want 
to elucidate the mechanism that has created the broad-lines despite the
supposedly still existing thick torus that blocks 
the inner nuclear region from the line of sight.
In particular it would be interesting to check if the evolution of 
the flare experienced  by  NGC~7582 shows a behaviour 
similar to SN~1988Z or NGC~5548 (common to other Seyfert~1), represented 
in Figure~2. This does not necessarily be the case if the 
variation is driven by a change in the column of obscuring material
that crosses our line of sight.

\section*{Acknowledgments}
We would like to thank the ESO-La Silla staff and particularly
J. Brewer and F. Patat for their diligent and efficient help,
and I.M. Hook who shared with us part of her allocated observing time.
We are indebted to G. Tenorio-Tagle
and  M. Mas-Hesse for very useful discussions and 
suggestions on an early draft of this paper.

\clearpage
 
\begin{deluxetable}{lllcrccc}
\footnotesize
\tablecaption{Journal of observations}
\tablewidth{0pt}
\tablehead{
\colhead{Date} & \colhead{Telescope}   & \colhead{Instrument} & 
\colhead{Slit}  & \colhead{Exp. t} & \colhead{Range} & 
\colhead{Resol.}     & \colhead{Seeing} \\
              &  &  & (arcsec)     & (s)  &   
(\AA)  & (\AA) & (arcsec)}
\startdata
1998 Jul 11     & Danish 1.54m & DFOSC + Loral/Lesser CCD & 2.0  
& 2700 & 3700--6700 & 3  & 1.5\nl
1998 Oct 6      & Danish 1.54m & DFOSC +  Loral/Lesser CCD & 1.5  
& 3600 & 3700--6700 & 3  & 1.5 \nl
1998 Oct 21     & ESO 3.6m  & EFOSC2 + Lesser CCD & 1.5 
& 900  & 3380--7540 & 4  & 1.3\nl
\enddata
\end{deluxetable}

\clearpage

\clearpage

\figcaption[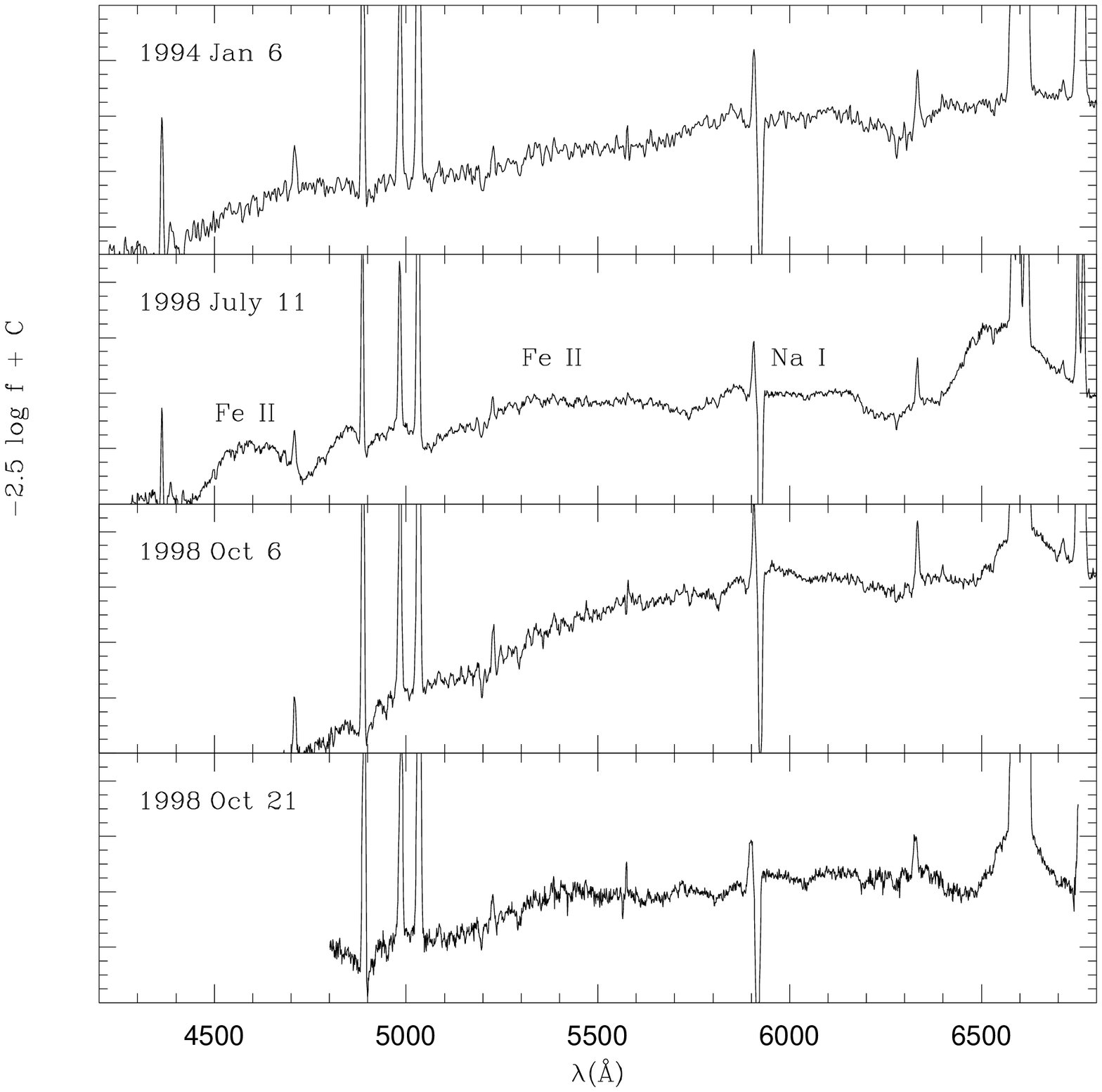]{Spectra of NGC~7582's nucleus}

\figcaption[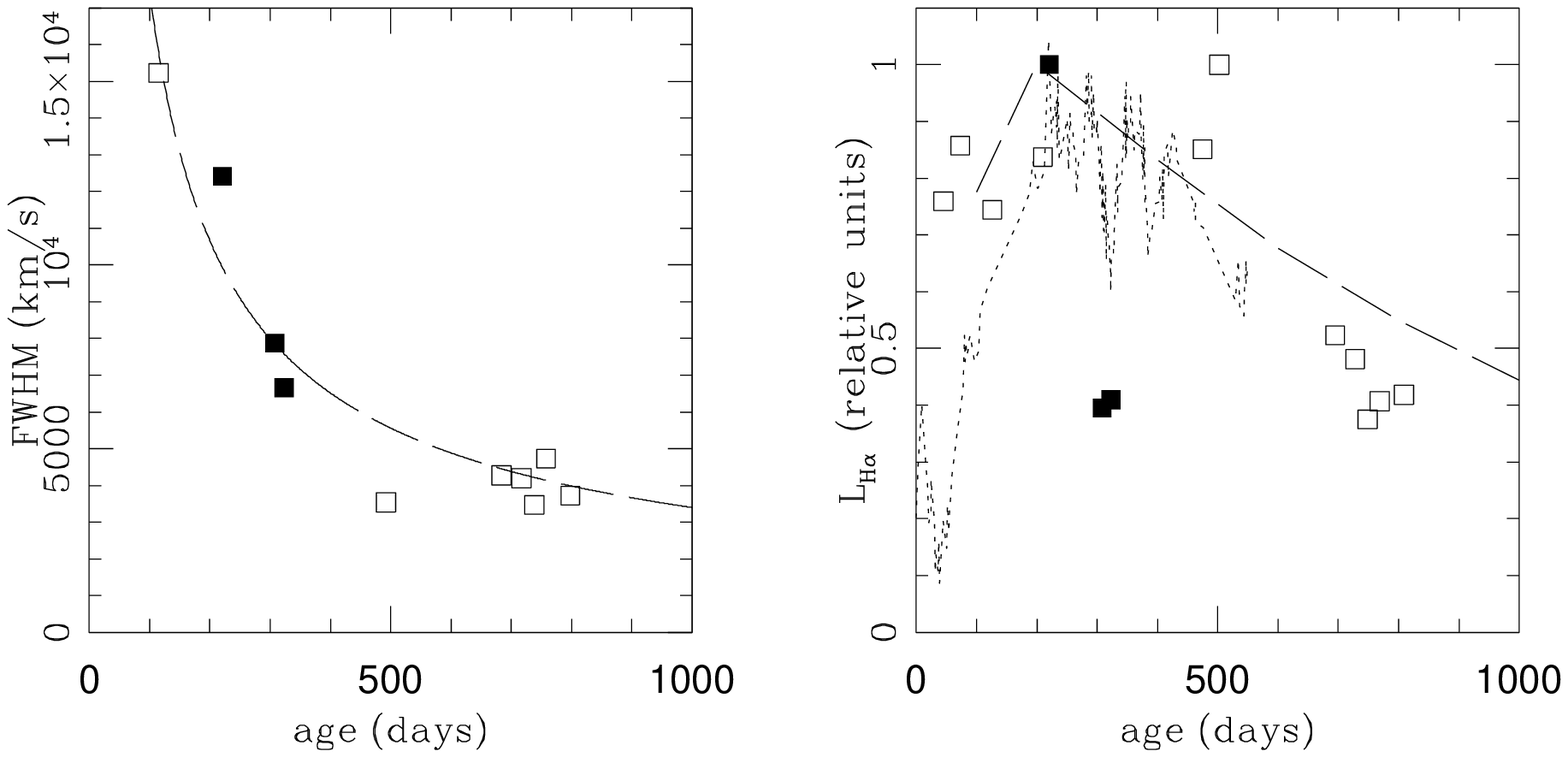]{
Evolution of \Ha\ line width and luminosity for NGC~7582
(filled squares), SN~1988Z (empty squares), NGC~5548 (dotted line)
and the semi-analytical solution for a cSNR  (dashed line)}



\clearpage

\plotone{specs.ps}

\clearpage

\plotone{Ha.ps}

\end{document}